\documentclass[3p]{elsarticle}

\usepackage{latexsym}
\usepackage{epsfig}
\usepackage{amssymb}
\usepackage{amsmath,bbm}
\usepackage{amsfonts,color}
\usepackage{amssymb,float}
\usepackage[utf8]{inputenc}
\usepackage{enumerate}

 \usepackage{hyperref}
\usepackage{enumitem}

\newcommand{\Lag}{{\mathcal L}}
\newcommand{\mS}{{\mathcal S}}
\newcommand{\mR}{{\mathcal R}}
\newcommand{\mD}{{\mathcal D}}
\newcommand{\mG}{{\mathcal G}}
\newcommand{\mP}{{\mathcal P}}
\newcommand{\mU}{{\mathcal U}}

\newcommand{\mpl}{M_{\rm Pl}}

\newcommand{\GG}{\mathbb{G}}

\newcommand{\GTGR}{\mathring{\mathbb{G}}}

\newcommand{\Qb}{\bar{Q}}

\newcommand{\cc}{\tilde{c}}

\newcommand{\be}{\begin{equation}}
\newcommand{\ee}{\end{equation}}

\newcommand{\bea}{\begin{eqnarray}}
\newcommand{\eea}{\end{eqnarray}}

\newcommand{\GL}{GL(4,\mathbb{R})}
\newcommand{\gl}{\mathfrak{gl}(4,\mathbb{R})}
\newcommand{\so}{\mathfrak{so}(3,1)}

\newcommand{\dd}{{\rm d}}

\begin{document}

\title{General Teleparallel Quadratic Gravity}
\author{Jose Beltr\'an Jim\'enez}
\address{Departamento de F\'isica Fundamental and IUFFyM, Universidad de Salamanca, E-37008 Salamanca, Spain.}
\ead{jose.beltran@usal.es}

\author{Lavinia Heisenberg} 
\address{Institute for Theoretical Physics,
ETH Zurich, Wolfgang-Pauli-Strasse 27, 8093, Zurich, Switzerland.}
\ead{lavinia.heisenberg@phys.ethz.ch}

\author{Damianos Iosifidis}
\address{Institute of Theoretical Physics, Department of Physics Aristotle University of Thessaloniki, 54124 Thessaloniki, Greece.}
\ead{diosifid@auth.gr}

\author{Alejandro Jim\'enez-Cano}
\address{Departamento de F\'isica Te\'orica y del Cosmos and CAFPE, Facultad de Ciencias, Avda Fuentenueva s/n, Universidad de Granada, 18071 Granada, Spain.}
\ead{alejandrojc@ugr.es}

\author{Tomi S. Koivisto}
\address{Laboratory of Theoretical Physics, Institute of Physics, University of Tartu, W. Ostwaldi 1, 50411 Tartu, Estonia.}
\address{National Institute of Chemical Physics and Biophysics,
R?avala pst. 10, 10143 Tallinn, Estonia.}
\ead{t.s.koivisto@astro.uio.no}

\date{\today}

\begin{abstract}
In this Letter we consider a general quadratic parity-preserving theory for a general flat connection. Imposing a local symmetry under the general linear group singles out the general teleparallel equivalent of General Relativity carrying both torsion and non-metricity. We provide a detailed discussion on the teleparallel equivalents of General Relativity and how the two known equivalents, formulated on Weitzenb\"ock and symmetric teleparallel geometries respectively, can be interpreted as two gauge-fixed versions of the general teleparallel equivalent. We then explore the viability of the general quadratic theory by studying the spectrum around Minkowski. The linear theory generally contains two symmetric rank-2 fields plus a 2-form and, consequently, extra gauge symmetries are required to obtain potentially viable theories. 
\end{abstract}

\date{\today}


\maketitle
\newpage

\section{Introduction}
One of the most beautiful properties of General Relativity (GR) is its intimate alliance with the geometry of spacetime. Nowadays it is understood that the geometrical interpretation of gravity arises as a consistency requirement for the low energy effective theory describing an interacting massless spin-2 particle. Since Einstein first taught us how to think of gravity in terms of the curvature of spacetime, we have become acquainted with this description which has proven to be extremely useful for studying gravitational phenomena as well as exploring possible modifications of gravity. 

However, the geometry of spacetime admits a much richer structure than that prescribed by GR once we unleash the affine sector. Remarkably, although rarely mentioned in standard textbooks, it is known that flat geometries with their well-defined notion of parallelism, provide alternative and fully equivalent representations of GR. On one hand, Weitzenb\"ock spaces can host a Teleparallel Equivalent of GR (TEGR) \cite{Teleparallel} where gravity is identified with torsion. On the other hand, flat and torsion-free spacetimes only containing a non-trivial non-metricity can also accommodate a Symmetric Teleparallel Equivalent of GR (STEGR) \cite{Nester:1998mp,BeltranJimenez:2017tkd}. In addition to the interest of these alternative formulations by themselves, they serve as different starting points to explore gravity theories beyond GR. The goal of this Letter is to extend previous studies in the literature on teleparallel geometries by allowing both torsion and non-metricity while keeping a trivial curvature, so the only constraint we impose is
\be
R^\alpha{}_{\beta\mu\nu}=2\partial_{[\mu}\Gamma^\alpha{}_{\nu]\beta}+2\Gamma^\alpha{}_{[\mu|\lambda|} \Gamma^\lambda{}_{\nu]\beta}=0.
\label{eq:zeroR}
\ee
This condition fixes the connection to be a pure $\GL$ gauge (also called {\it inertial connection}) so that it can be expressed in terms of an arbitrary $\Lambda^\alpha{}_\beta\in\GL$ as
\be
\Gamma^\alpha{}_{\mu\beta}=(\Lambda^{-1})^\alpha{}_\rho\partial_\mu\Lambda^\rho{}_\beta.
\label{eq:Telecon}
\ee
This inertial connection features a global symmetry $\Lambda\rightarrow \mU\Lambda$ for a constant $\mU\in\GL$ that will be present in the teleparallel theories. The torsion and the non-metricity of the teleparallel geometry are given by
\begin{align}
T^\alpha{}_{\mu\beta}=&2\Gamma^\alpha{}_{[\mu\beta]}=2(\Lambda^{-1})^\alpha{}_\rho\partial_{[\mu}\Lambda^\rho{}_{\beta]},
\label{eq:Teletorsion}\\
Q_{\alpha\mu\nu}=&\nabla_\alpha g_{\mu\nu}=\partial_\alpha g_{\mu\nu}-2(\Lambda^{-1})^\lambda{}_\rho\partial_\alpha\Lambda^\rho{}_{(\mu} g_{\nu)\lambda},
\end{align}
respectively, with the independent traces $T_\mu=T^\alpha{}_{\mu\alpha}$, $Q_\mu=Q_{\mu\alpha}{}^\alpha$ and $\Qb_\mu=Q^{\alpha}{}_{\alpha\mu}$. These are the two fundamental objects to construct a general teleparallel theory. Notice that, since these objects transform covariantly under Diffeomorphisms (Diffs), the resulting theory will automatically enjoy this symmetry (unless it is explicitly broken from the outset). It may be worth to emphasise that $\Lambda$ is not a tensor under Diffs, as can be easily seen from the transformation of $\Gamma$ as a connection. The action will be the most general quadratic and parity-preserving form built in terms of these objects. We parameterise it as
\begin{align}\label{eqSIIgen}	
\mathcal{S}_{\parallel} =\frac12\mpl^2\int\dd^4x\sqrt{-g}\Big[& a_{1}T_{\alpha\mu\nu}T^{\alpha\mu\nu} +
	a_{2}T_{\alpha\mu\nu}T^{\nu\mu\alpha} +
	a_{3}T_{\mu}T^{\mu}+ b_{1}Q_{\alpha\mu\nu}T^{\nu\alpha\mu}+
	b_{2}Q_{\mu}T^{\mu} +
	b_{3}\Qb_{\mu}T^{\mu}    \nonumber\\
&+c_{1}Q_{\alpha\mu\nu}Q^{\mu\alpha\nu} +
	c_{2}Q_{\alpha\mu\nu}Q^{\mu\nu\alpha} +
	c_{3}Q_{\mu}Q^{\mu}+
	c_{4}\Qb_{\mu}\Qb^{\mu}+
	c_{5}Q_{\mu}\Qb^{\mu}\Big].
\end{align}
We will use the shortcut $\mathcal{S}_{\parallel} =\frac12\mpl^2\int\dd^4x\sqrt{-g}\,\GG$ with $\GG$ implicitly defined by \eqref{eqSIIgen}. This action reduces to New GR \cite{Hayashi:1979qx} for a metric connection with $Q_{\alpha\mu\nu}=0$ and to Newer GR \cite{BeltranJimenez:2017tkd} for a torsion-free connection. The corresponding field equations are obtained by varying w.r.t the fundamental fields $g_{\mu\nu}$ and $\Lambda^\alpha{}_\beta$. The metric field equations can then be expressed as
\be
\mG_{\mu\nu}=\frac{1}{\mpl^2} T_{\mu\nu}
\ee
where we have defined
\begin{align}
\mG^{\mu\nu}\equiv\frac{2\mpl^{-2}}{\sqrt{-g}}\frac{\delta \mathcal{S}_{\parallel} }{\delta g_{\mu\nu}} 
  & =\frac12\GG g^{\mu\nu} + a_1\Big(T^{\mu}{}_{\sigma\rho}T^{\nu\sigma\rho} - 2T^{\rho\sigma\mu}T_{\rho\sigma}{}^{\nu}\Big) - a_2 T^{\rho\sigma\mu}T_{\sigma\rho}{}^{\nu} - a_3 T^{\mu}T^{\nu}\nonumber \\
  & \quad - b_1 T^{\rho\sigma(\mu}\Big(Q^{\nu)}{}_{\sigma\rho} -Q_{\sigma\rho}{}^{\nu)}\Big) + b_2 \Big(Q^{\rho\mu\nu}T_{\rho} + Q^{(\mu}T^{\nu)}\Big) + b_3 \Big(Q^{(\mu\nu)\rho}T_\rho + \Qb^{(\mu}T^{\nu)}\Big)\nonumber \\
  & \quad - c_1 \Big(Q^\mu{}_{\sigma\rho}Q^{\nu\sigma\rho}+2Q^{\sigma\rho\mu}Q_{\sigma\rho}{}^\nu\Big)-c_2 \Big(2Q_{\sigma\rho}{}^{(\mu}Q^{\nu)\sigma\rho}+Q^{\rho\sigma\mu}Q_{\sigma\rho}{}^\nu\Big)-c_3 \Big(Q^{\mu}Q^{\nu}+2Q_{\rho}Q^{\rho\mu\nu}\Big)\nonumber \\
  & \quad - c_4 \Big(\Qb^\mu \Qb^\nu+2\Qb_\rho Q^{(\mu\nu)\rho}\Big)-c_5 \Big(Q^{(\mu}\Qb^{\nu)}+\Qb_\rho Q^{\rho\mu\nu}+Q_\rho Q^{(\mu\nu)\rho}\Big)\nonumber \\
  & \quad - \left(\nabla_\rho +\frac{1}{2}Q_\rho +T_{\rho}\right) \Big[2c_1Q^{\rho\mu\nu}+2c_2 Q^{(\mu\nu)\rho}+2c_3 Q^{\rho}g^{\mu\nu}+2c_4 g^{\rho(\mu}\Qb^{\nu)}\nonumber \\
  & \qquad\qquad\qquad+c_5 \Big(\Qb^\rho g^{\mu\nu}+ g^{\rho(\mu}Q^{\nu)}\Big)-b_1T^{(\mu\nu)\rho}+b_2 T^\rho g^{\mu\nu}+b_3 g^{\rho(\mu}T^{\nu)}\Big]\,,
\end{align}
and the energy-momentum tensor of the matter sector $\mS_{\rm m}$ is given by the usual expression
\be
T^{\mu\nu}=-\frac{2}{\sqrt{-g}}\frac{\delta\mS_{\rm m}}{\delta g_{\mu\nu}}.
\ee
To compute the field equations for $\Lambda^\alpha{}_\beta$, we use the identity for general connections that gives the variation of the connection under an infinitesimal gauge transformation parameterised by $\epsilon$ in terms of the corresponding covariant derivative of the gauge parameter $\delta_\epsilon\Gamma=\nabla\epsilon$. Applied to our teleparallel connection and taking into account that it is pure gauge, we then obtain that\footnote{It may be convenient to be a little more explicit for clarity. A given connection $\Gamma$ transforms under the corresponding gauge transformation parameterised by $\mU$ as  $\Gamma\rightarrow \mU^{-1}(\Gamma+\dd)\mU$. If the connection is pure gauge $\Gamma=\Lambda^{-1}\dd\Lambda$, this transformation simply leads to the expected transformation $\Lambda\rightarrow\Lambda\mU$. For an infinitesimal transformation $\mU=\mathbbm{1}+\epsilon$, we have $\delta_\epsilon\Gamma=\dd\epsilon+[\Gamma,\epsilon]=\nabla\epsilon$. Since the pure gauge connection changes under the infinitesimal transformation as $\Lambda\rightarrow\Lambda+\Lambda\epsilon$, we can relate it to a variation of $\Lambda$ with $\epsilon=\Lambda^{-1}\delta\Lambda$ that leads to \eqref{eq:deltaGamma}.}
\be
\delta\Gamma^\alpha{}_{\mu\beta}=\nabla_\mu\big[(\Lambda^{-1})^\alpha{}_\rho\delta\Lambda^\rho{}_\beta\big].
\label{eq:deltaGamma}
\ee
The field equations can then be expressed as
\be
\big(\nabla_\mu+T_\mu\big)\mP_\alpha{}^{\mu\nu}=0
\ee
where we have defined
\begin{align}
\mP_\rho{}^{\mu\nu}\equiv\frac{\delta \mathcal{S}_{\parallel} }{\delta\Gamma^\rho{}_{\mu\nu}} 
  & =\sqrt{-g} \mpl^{2}\Big[ 2a_1 T_\rho{}^{\mu\nu}- 2 a_2 T^{[\mu\nu]}{}_\rho + 2a_3 T^{[\mu}\delta_{\rho}^{\nu]}\nonumber \\
  & \quad + b_1 \Big(Q^{[\mu\nu]}{}_\rho + T^{(\lambda\nu)\mu} g_{\lambda\rho}\Big) + b_2 \Big(Q^{[\mu}\delta_\rho^{\nu]}-T^{\mu}\delta_\rho^\nu\Big) + b_3 \Big(\Qb^{[\mu} \delta_\rho^{\nu]}-g^{\mu(\nu}T^{\lambda)} g_{\lambda\rho}\Big) \nonumber \\
  & \quad - 2c_1 Q^{\mu\nu}{}_{\rho} - 2c_2 Q^{(\nu\lambda)\mu}g_{\lambda\rho} - 2c_3 Q^{\mu}\delta_{\rho}^{\nu} - 2c_4 g^{\mu(\nu}\Qb^{\lambda)}g_{\lambda\rho} - c_5 \Big(\Qb^\mu 4\delta_{\rho}^{\nu}+g^{\mu(\nu}Q^{\lambda)}g_{\lambda\rho}\Big)\Big]\,.
\end{align}
We have assumed that the connection does not enter the matter action so the hypermomentum vanishes. We could have used a Palatini approach by allowing a fully general affine connection and imposing the constraint \eqref{eq:zeroR} with suitable Lagrange multipliers. We however prefer to solve the constraint and formulate the theory in terms of the fundamental fields, namely  $g_{\mu\nu}$ and $\Lambda^\alpha{}_\beta$ directly so we do not have to solve for any Lagrange multipliers. The fundamental fields make up a total of $10+16=26$ independent components. However, the gauge symmetry provided by Diffs invariance reduce these to a maximum of 18 propagating fields that can be associated with the 16 components of $\Lambda^\alpha{}_\beta$ plus the two polarisations of the graviton contained in $g_{\mu\nu}$. These are still too many dof's to avoid ghostly modes so we need to restrict the parameter space to render the theory stable. But before moving to that, let us delve into the theories with a trivial $\Lambda$-sector that actually reduce to GR.

\section{On the equivalents of GR}
It is known that the teleparallel framework permits alternative formulations of GR, namely TEGR and STEGR. Here we want to elaborate further on the existence of these equivalences and clarify their origins. The starting point is the known post-Riemannian expansion of the Ricci scalar when the general connection is expanded around the Levi-Civita connection of the spacetime metric as $\Gamma^\alpha{}_{\mu\nu}=\{^\alpha{}_{\mu\nu}\}+\Omega^\alpha{}_{\mu\nu}$. In that case, the Ricci scalar can be expressed as
\be
R=\mR+ 2g^{\mu\nu}\left(\mD_{[\alpha}\Omega^\alpha{}_{\mu]\nu} +\Omega^\alpha{}_{[\alpha\lvert\lambda\rvert}\Omega^\lambda{}_{\mu]\nu}\right)\, ,
\label{eq:RtoROmega}
\ee
with $\mR$ and $\mD$ the Ricci scalar and covariant derivative of the Levi-Civita connection respectively. We can further decompose $\Omega^\alpha{}_{\mu\nu}=L^\alpha{}_{\mu\nu}+K^\alpha{}_{\mu\nu}$ in terms of the disformation and the contorsion tensors defined as $
L^\alpha_{\phantom{\alpha}\mu\nu}  = \frac{1}{2} Q^{\alpha}_{\phantom{\alpha}\mu\nu} - Q_{(\mu\phantom{\alpha}\nu)}^{\phantom{(\mu}\alpha}$ and $K^\alpha_{\phantom{\alpha}\mu\nu} = \frac{1}{2}T^\alpha_{\phantom{\alpha}\mu\nu} + T_{(\mu{\phantom{\alpha}\nu)}}^{\phantom{,\mu}\alpha}$.
 In terms of these objects, we can write \eqref{eq:RtoROmega} as
\be
R=\mR+\GTGR+\mD_{\mu}\left(Q^{\mu}-\Qb^{\mu}+2 T^{\mu}\right)
\label{eq:RtoROmega2}
\ee
where we have defined
\begin{align}
\GTGR=&\frac{1}{4} T_{\mu \nu \rho} T^{\mu \nu \rho}+\frac{1}{2} T_{\mu \nu \rho} T^{\mu \rho \nu}-T_{\mu} T^{\mu}+Q_{\mu \nu \rho} T^{\rho\mu \nu}-Q_{\mu} T^{\mu}+\bar{Q}_{\mu} T^{\mu}\nonumber\\
&+\frac{1}{4} Q_{\mu \nu \rho} Q^{\mu \nu \rho}-\frac{1}{2} Q_{\mu \nu \rho} Q^{\nu \mu \rho}-\frac{1}{4} Q_{\mu} Q^{\mu}+\frac{1}{2} Q_{\mu} \bar{Q}^{\mu}
\end{align}
obtained from $\GG$ upon the parameter choice \cite{Iosifidis:2018zwo,Iosifidis:2019jgi}
\be
(a_1,a_2,a_3)=\left(\frac14,\frac12,-1\right),\quad (b_1,b_2,b_3)=(1,-1,-1)\quad \text{and}\quad (c_1,c_2,c_3,c_4,c_5)=\left(\frac14,-\frac12,-\frac14,0,\frac12\right). 
\ee
The general relation \eqref{eq:RtoROmega2} between the Ricci scalars, barring the irrelevant total derivative, is the root for the equivalents of GR in teleparallel geometries. As a matter of fact, this very possibility can be further traced back to the equivalence between the metric and the Palatini formalisms for the Einstein-Hilbert action. The field equation for $\Omega^\alpha{}_{\beta\gamma}$ of the action $\mS=\frac12\mpl^2\int\dd^4x\sqrt{-g} R$ built in terms of \eqref{eq:RtoROmega} is
\be
g^{\rho\sigma}\Omega^\gamma{}_{\rho\sigma}\delta^\beta_\alpha+g^{\beta\gamma}\Omega^\rho{}_{\rho\alpha}-g^{\beta\rho}\Omega^\gamma{}_{\alpha\rho}-g^{\rho\gamma}\Omega^\beta{}_{\rho\alpha}=0
\ee
that sets a vanishing $\Omega^\alpha{}_{\mu\beta}$ up to the non-trivial kernel of the above equation spanned by a projective mode $\Omega^\alpha{}_{\beta\gamma}=A_\beta\delta^\alpha_\gamma$. Therefore, the general solution for the connection is $\Gamma^\alpha{}_{\mu\nu}=\{^\alpha{}_{\mu\nu}\}+A_\mu\delta^\alpha_\nu$. 
This clearly shows that the geometry of GR admits a whole class of projective geometries parameterised by $A_\mu$ with torsion $T^\alpha{}_{\mu\nu}=2A_{[\mu}\delta^\alpha_{\nu]}$ and a Weyl non-metricity $Q_{\alpha\mu\nu}=A_\alpha g_{\mu\nu}$. By suitably fixing the projective gauge mode we can choose how much torsion and/or non-metricity we want. This of course lacks any physical relevance and can only be returned to having any physical interest via couplings in the matter sector. In the usual case with minimally coupled fields respecting the projective symmetry, the issue remains completely irrelevant.  

Let us now turn to the case of our flat connections\footnote{Since the Riemann tensor is invariant under projective transformations with $A_\mu=\partial_\mu A$, the flat condition does not fully fix the projective symmetry. For instance, the Weyl non-metricity trace in a teleparallel geometry reduces to a Weyl integrable pure gradient $Q_\mu=\partial_\mu\log\frac{\det g}{(\det \Lambda)^{2}}$.} for which we have
\be 
\mR=-\GTGR+\mathcal{D}_{\mu}\left(Q^{\mu}-\bar{Q}^{\mu}+2 T^{\mu}\right). 
\ee
This shows that the two actions
\begin{align}
\mS_{\rm EH}[g]=-\frac12\mpl^2\int\dd^4x\sqrt{-g}\mR(g)\quad{\rm and}\quad \mS_{{\rm GR}_\parallel}[g,\Lambda]=\frac12\mpl^2\int\dd^4x\sqrt{-g}\;\GTGR
\end{align}
describe the same theory, up to the boundary term. The above action $\mS_{{\rm GR}_\parallel}[g,\Lambda]$ constitutes the General Teleparallel Equivalent of GR (GTEGR). It may not be obvious that the geometrical identity used will lead to the same dynamics, with a fair objection being that the number of fields on both sides is not the same. While we need only the metric to construct the Einstein-Hilbert action, the action $\mS_{{\rm GR}_\parallel}[g,\Lambda]$ contains the metric and the connection parameterised by $\Lambda$. Thus, we need some additional symmetries in order to square the number of dof's in both sides. This can be easily shown by noticing that $
\delta_\Lambda \mS_{{\rm GR}_\parallel}[g,\Lambda]=0 $
is satisfied off-shell, which means that $\Lambda$ can at most contribute a total derivative to the action. This can be shown in a very transparent and straightforward manner by going to the vierbein formulation of the theory. The Einstein-Hilbert Lagrangian in that formalism is simply $\Lag_{\rm EH}=\epsilon_{abcd} R^{ab}\wedge e^c\wedge e^d$, where $R^a{}_b=\dd \omega^a{}_b+\omega^a{}_m\wedge \omega^m{}_b$ is the curvature 2-form of the connection $\omega^a{}_b$ and $\epsilon_{abcd}$ is the Levi-Civita tensor. We can consider the post-Riemannian expansion of the connection $\omega=\bar{\omega}+\Omega$, with $\bar{\omega}$ the metric-compatible and torsion-free Levi-Civita part. The Einstein-Hilbert Lagrangian in this decomposition takes the form 
\be
 \Lag_{\rm EH}=\epsilon_{abcd} \left(\bar{R}^{ab}+\bar{D} \Omega^{ab}+\Omega^a{}_m\wedge\Omega^{mb}\right)\wedge e^c\wedge e^d
 \label{eq:RToRbvierbein}
\ee
that is the vierbein equivalent of \eqref{eq:RtoROmega}. Again, we can drop the total derivative term involving $\bar{D} \Omega^{ab}$ that can be written as $\dd\big(\epsilon_{abcd}\Omega^{ab}\wedge e^c\wedge e^d\big)$ because the Levi-Civita connection is metric compatible, so $\bar{D}\epsilon_{abcd}=\bar{D}\eta_{ab}=0$, and torsion-free, so  $\bar{D}e^a=0$. If we now impose the flatness condition on the full curvature $R^a{}_b=0$ we recover the relation $\bar{R}^a{}_b=-(\bar{D}\Omega^a{}_b+\Omega^a{}_m\wedge\Omega^m{}_b)$ and the connection is $\omega=\bar{\omega}+\Omega=\Lambda^{-1}\dd\Lambda$ so we have $\Omega=-\bar{\omega}+\Lambda^{-1}\dd\Lambda$. If we plug this into the $\Omega\wedge\Omega$ term in \eqref{eq:RToRbvierbein} we can write
\be
\epsilon_{abcd}\Omega^a{}_m\wedge\Omega^{mb}\wedge e^c\wedge e^d=\epsilon_{abcd}\left[\bar{\omega}^a{}_m\wedge\bar{\omega}^{mb}-\bar{D}\left(\Lambda^{-1}\dd\Lambda\right)^{ab}\right]\wedge e^c\wedge e^d\,.
\ee
We clearly see that $\Lambda$ only contributes a total derivative and we obtain the Einstein Lagrangian\footnote{A word on nomenclature: We call the Einstein Lagrangian/action to the dynamical part of the Einstein-Hilbert action without the total derivative necessary to conform a scalar, i.e., $\Lag_{\rm Einstein}=\frac12\mpl^2\sqrt{-g}g^{\mu\nu}\Big(\left\{^{\phantom{i} \alpha}_{\beta\mu}\right\} \left\{^{\phantom{i} \beta}_{\nu\alpha}\right\} -\left\{^{\phantom{i} \alpha}_{\beta\alpha}\right\}\left\{^{\phantom{i} \beta}_{\mu\nu}\right\} \Big).$ \label{EinsteinLag}} of GR as the only dynamical part. Consequently, the action built in terms of $\GTGR$ is nothing but the special quadratic theory where the global $\GL$ invariance becomes a local symmetry. In other words, we can perform a local $\GL$ transformation to fully trivialise the connection. In that gauge, we have $\Omega^\alpha{}_{\mu\nu}=-\{^\alpha{}_{\mu\nu}\}$ and a vanishing torsion, while $Q_{\alpha\mu\nu}=\partial_\alpha g_{\mu\nu}$ so that $\mS_{{\rm GR}_\parallel}$ reduces to the Einstein action of GR, thus providing the $\GL$ generalisation of the coincident gauge in STEGR \cite{BeltranJimenez:2017tkd}.

In view of GTEGR, we can give a nice interpretation of TEGR and STEGR as different gauge-fixed versions of the general equivalent:

\begin{itemize}
\item {\bf TEGR gauge}. 

In TEGR, the connection is further restricted to be metric compatible. The corresponding additional constraint $\nabla_\alpha g_{\mu\nu}=0$, for the teleparallel connection, leads to the equation
\be
2(\Lambda^{-1})^\lambda{}_\kappa\partial_\alpha\Lambda^\kappa{}_{(\mu} g_{\nu)\lambda} =\partial_\alpha g_{\mu\nu}\,. 
\label{eq:Lambdag}
\ee
that relates the metric and $\Lambda$. This gauge does not fix the full $\GL$ symmetry, but there is still some residual symmetry. To reveal the unfixed sector of the symmetry, we notice that the above equation is solved by a teleparallel connection satisfying
\be
g_{\mu\nu}=\Lambda^\alpha{}_\mu\Lambda^\beta{}_\nu c_{\alpha\beta}
\ee
for an arbitrary constant $c_{\alpha\beta}$. It is thus clear that the gauge-fixing \eqref{eq:Lambdag} leaves undetermined the orthogonal subgroup with respect to the metric $c_{\alpha\beta}$. Since we are interested in Lorentzian metrics, it is natural to choose $c_{\alpha\beta}=\eta_{\alpha\beta}$ so that the residual symmetry is nothing but a local Lorentz invariance, which is the well-known symmetry of TEGR.

\item {\bf STEGR gauge}.

The STEGR on the other hand is obtained by imposing  $T^\alpha{}_{\mu\nu}=0$. This gauge constrains the teleparallel connection to have the form $\Lambda^\alpha{}_\beta=\partial_\beta\xi^\alpha$ for some arbitray $\xi^\alpha$'s that can be identified with a coordinate transformation. This gauge represents a minimal covariantisation of GR in the sense that is the maximal partial fixing of the local $\GL$ that still allows a covariant formulation. In fact, the parameters $\xi^\alpha$ that parameterise the local symmetry in this partial gauge fixing can be nicely interpreted as St\"uckelberg fields introduced to restore covariance in the Einstein Lagrangian.
\end{itemize}

The above are special gauges because only one of the fundamental geometrical objects of the affine structure is left and, thus, gravity is fully described in terms of them. Together with the usual description of GR as the spacetime curvature, this tern has been dubbed the geometrical trinity \cite{BeltranJimenez:2019tjy}. However, our discussion clarifies that it is possible to make other gauge choices within the teleparallel realm that interpolate between these two special gauges.

The identification of TEGR and STEGR as merely particular cases of GTEGR with different gauge-fixing terms opens up the possibility for a whole plethora of modifications of gravity  based on non-linear extensions of the corresponding partially gauge-fixed version of the GR equivalent analogous to the $f(T)$ \cite{Bengochea:2008gz,Li:2010cg} and $f(Q)$ \cite{Jimenez:2019ovq} theories based on the TEGR and STEGR gauges. It is now apparent that the differences between these two extensions root in the different gauges employed for their construction. In general, the different non-linear extensions can be parameterised as
\be
\mS=\int\dd^4x\Big[\sqrt{-g}f(\GTGR)+F(\Lambda,\lambda,\cdots)\Big]
\ee
where $F$ is the gauge-fixing condition that depends on the connection $\Lambda$, some Lagrange multipliers $\lambda$ and, possibly, other fields such as the metric. The non-linear extensions based on TEGR and STEGR are those with $F_{\rm TEGR}=\lambda^{\alpha\mu\nu}\nabla_\alpha g_{\mu\nu}$ and $F_{\rm STEGR}=\lambda_\alpha{}^{\mu\nu} T^\alpha{}_{\mu\nu}$ respectively.

 It is important to strongly emphasise however that most, if not all, of those extensions will be prone to suffer from some pathologies owed to the loss of symmetries. This is  in particular the case of $f(T)$ \cite{Li:2010cg} and $f(Q)$ \cite{Jimenez:2019ovq} theories. Any cosmological solution is strongly coupled in the former, while the latter alleviates the situation by only suffering from strong coupling on maximally symmetric backgrounds. We notice that fully fixing the gauge that trivialises the connection, i.e., $F=\lambda_\alpha{}^\beta(\Lambda^\alpha{}_\beta-\delta^\alpha{}_\beta)$, leads to the Coincident GR so the corresponding extensions will be the same as $f(Q)$. A potentially interesting non-linear extension could be that without any partial gauge-fixing, i.e., $f(\GTGR)$ where the full $\Lambda^\alpha{}_\beta$ is allowed to contribute.

\section{Perturbative spectrum on Minkowski}
\subsection{Quadratic Lagrangian}
As the first step towards unveiling the viable Lagrangians we will study the linear theory on a Minkowski background. For that, we will expand the connection and the metric to first order as follows
\bea
g_{\mu\nu}=\eta_{\mu\nu}+h_{\mu\nu},\quad\quad
\Lambda^\alpha{}_\beta=\delta^\alpha{}_\beta+\lambda^\alpha{}_\beta.
\eea
The torsion and the non-metricity are given by:
\bea
T^\alpha{}_{\mu\nu}=2\partial_{[\mu}\lambda^\alpha{}_{\nu]},\quad\quad
Q_{\alpha\mu\nu}=\partial_\alpha \big(h_{\mu\nu}-H_{\mu\nu}\big),
\label{eq:pertQ}
\eea
where we have defined $H_{\mu\nu}=2\lambda_{(\mu\nu)}$. We will also introduce the antisymmetric part of the connection perturbation as $B_{\mu\nu}=2\lambda_{[\mu\nu]}$ so that we have $\lambda_{\mu\nu}=\frac12(H_{\mu\nu}+B_{\mu\nu})$. The quadratic Lagrangian in terms of these fields reads
\begin{align}\label{quadrSinHhB}
\frac{1}{\mpl^2}\Lag_{\parallel}^{(2)}=&\;\;\;\;\frac{c_1}{2}\partial_\alpha h_{\mu\nu} \partial^\alpha h^{\mu\nu}+\frac{c_{24}}{2}\partial_\mu h^{\mu\alpha}\partial^\nu h_{\nu\alpha}+\frac{c_5}{2}\partial_\mu h\partial_\nu h^{\mu\nu}+\frac{c_3}{2} (\partial h)^2\nonumber\\
&+\frac{\cc_1}{8}\partial_\alpha H_{\mu\nu} \partial^\alpha H^{\mu\nu}+\frac{\cc_2}{8}\partial_\mu H^{\mu\alpha}\partial^\nu H_{\nu\alpha}+\frac{\cc_3}{4}\partial_\mu H\partial_\nu H^{\mu\nu}+\frac{\cc_4}{8} (\partial H)^2\nonumber\\
&+\frac{2a_1-a_2}{8}\partial_\mu B_{\alpha\beta}\partial^\mu B^{\alpha\beta}-\frac{2a_1-3a_2-a_3}{8}\partial_\mu B^{\mu\alpha}\partial^\nu B_{\nu\alpha}\nonumber\\
&+\frac{b_1-4c_1}{4}\partial_\alpha h_{\mu\nu} \partial^\alpha H^{\mu\nu}-\frac{b_1+b_3+4c_{24}}{4}\partial_\mu h^{\mu\alpha}\partial^\nu H_{\nu\alpha}+\frac{b_3-2c_5}{4}\partial_\mu H\partial_\nu h^{\mu\nu}-\frac{b_2+2c_5}{4}\partial_\mu h\partial_\nu H^{\mu\nu}\nonumber\\
&+\frac{b_2-4c_3}{4} \partial_\mu h\partial^\mu H+\frac{2a_1+a_2+a_3-b_1+b_3}{4}\partial_\mu B^{\mu\alpha}\partial^\nu H_{\nu\alpha}+\frac{b_1-b_3}{4}\partial_\mu B^{\mu\alpha}\partial^\nu h_{\nu\alpha},
\end{align}
where we have defined
\begin{align}
&c_{24}=c_2+c_4,\quad\cc_1=2a_1+a_2-2b_1+4c_1,\quad\cc_2=-2a_1-a_2+a_3+2(b_1+b_3)+4c_{24},\nonumber\\
&\cc_3=-a_3+b_2-b_3+2c_5,\quad\cc_4=a_3-2b_2+4c_3.
\end{align}
The first three lines of \eqref{quadrSinHhB} describe the pure $H_{\mu\nu}$, $h_{\mu\nu}$ and $B_{\mu\nu}$ sectors respectively, while the last two lines comprise the mixings among the different fields. Notice that the parameters $c_2$ and $c_4$ are degenerate at linear order, since they only enter through the combination $c_{24}$. This degeneracy will be broken by interactions. For arbitrary parameters, this quadratic Lagrangian contains the 2-symmetric rank-2 fields $h_{\mu\nu}$ and $H_{\mu\nu}$ plus the antisymmetric field $B_{\mu\nu}$.  The absence of any masses for these fields is guaranteed by the $\GL$ global symmetry. In view of the fields that enter the quadratic Lagrangian, a very reasonable requirement in order to avoid ghost-like fields around Minkowski is that the field content should correspond to two massless spin-2 fields plus a massless Kalb-Ramond field, each with its corresponding gauge symmetry. Since the general quadratic theory only enjoys the usual Diff symmetry, the absence of unstable modes makes it necessary to fix some coefficients to enhance the symmetries.

\subsection{Symmetries}
It will be useful to analyse how the different symmetries of the theory are realised in the perturbations of the metric and the connection. By construction, we have Diffs that, at linear order, are realised as usual in the metric and the connection perturbations
\bea
\delta_\zeta h_{\mu\nu}=-2\partial_{(\mu}\zeta_{\nu)},\quad\quad
\delta_\zeta \lambda^\alpha{}_\beta=-\partial_\beta\zeta^\alpha.
\label{eq:Diff1}
\eea
The transformation under Diffs for the connection translates into 
\be
\delta_\zeta H_{\mu\nu}=-2\partial_{(\mu}\zeta_{\nu)}\quad \text{and}\quad \delta_\zeta B_{\mu\nu}=2\partial_{[\mu}\zeta_{\nu]}.
\label{eq:Diff2}
\ee
It will be important to notice that, since the background values of both the torsion and the non-metricity vanish, their perturbations are gauge-invariant under linearised Diffs, in compliance with the Stewart-Walker lemma.  On the other hand, the global symmetry enjoyed by the connection will be realised as $\delta_\omega\lambda^\alpha{}_\beta=\omega^\alpha{}_\beta$
with $\omega^\alpha{}_\beta$ a constant element of the Lie algebra $\gl$ so that we have
\bea
\delta_\omega H_{\mu\nu}=2\omega_{(\mu\nu)},\quad\quad
\delta_\omega B_{\mu\nu}=2\omega_{[\mu\nu]}.
\eea
Since $\omega$ is constant, this is a shift symmetry that will forbid any mass terms for both fields. Furthermore, the field $B_{\mu\nu}$ transforms with the component along $\so$ and so it takes care of the whole change under Lorentz transformations, while the symmetric field $H_{\mu\nu}$ will change with the complementary components in $\gl$. One important consequence of this is that the realisation of a local Lorentz symmetry will be straightforwardly identified in the quadratic Lagrangian from the absence of $B_{\mu\nu}$.

\subsection{Minimal field content: GTEGR}
An instructive case to look at is the general equivalent of GR introduced above. The quadratic Lagrangian for those parameters reduces to
\be
\frac{1}{\mpl^2}\Lag_{\rm GTEGR}=\frac18\partial_\alpha h_{\mu\nu} \partial^\alpha h^{\mu\nu}-\frac14\partial_\mu h^{\mu\alpha}\partial^\nu h_{\nu\alpha}+\frac14\partial_\mu h\partial_\nu h^{\mu\nu}-\frac18 (\partial h)^2.
\ee
We see here that only the metric perturbation contributes and precisely in the Fierz-Pauli form that guarantees the propagation of a healthy massless spin-2 particle, while all the dof's associated to the connection only contribute a total derivative, thus corroborating at linear level the realisation of the local gauge symmetry $\GL$ up to the total derivative term that we have dropped.  One may wonder whether, at linear order, there is a broader class of theories realising the linearised $\GL$ gauge symmetry, even if it is not realised at the non-linear order. This would diagnose a discontinuity in the number of dof's so we could already rule them out on the basis of phenomenological viability caused by a strong coupling problem on Minkowski. In order to analyse it, we can obtain the general linear theory for which the connection dof's $H_{\mu\nu}$ and $B_{\mu\nu}$ have trivial linear field equations. Doing so we obtain that this is only possible precisely for the parameters of GTEGR, up to a normalisation factor and the explained degeneracy between $c_2$ and $c_4$. Thus, requiring the full global symmetry to become a local symmetry, even at linear order, is a very strong condition that fixes the theory to be GR at that order (as should be) and only leaves one free parameter in the full theory. This is the minimal field content we can have if we want the theory to describe gravity. Theories with other parameters, again assuming that gravity is a required sector, will then propagate more dof's.

\subsection{Maximal field content}
The general quadratic Lagrangian contains the fields $h_{\mu\nu}$, $H_{\mu\nu}$ and $B_{\mu\nu}$ which are prone to propagate ghost-like modes unless appropriate gauge symmetries are present. Thus, the maximum number of physical dof's that we can have without incurring in ghostly instabilities will correspond to having 2 massless spin-2 fields plus a massless Kalb-Ramond field, adding up to a total of 2+2+1=5 dof's. Any choice of parameters propagating more than 5 dof's will have ghosts around a Minkowski background. It will then be useful to study those theories that precisely propagate this number of dof's. For the antisymmetric field $B_{\mu\nu}$ we will impose the usual gauge symmetry for a massless 2-form field given by $\delta B_{\mu\nu}=2\partial_{[\mu}\theta_{\nu]}$ for an arbitrary $\theta_\nu$. This gauge symmetry leads to the Bianchi identity
\be
\partial_\mu\left(\frac{\delta \mS^{(2)}_\parallel}{\delta B_{\mu\nu}}\right)=0
\ee
that imposes 
\be
\text{Gauge 2-form:}\quad\quad\quad2a_1+a_2+a_3=0 \quad \text{and}\quad b_3=b_1. 
\label{eq:gaugeB}
\ee
Since $B_{\mu\nu}$ does not contribute to $Q_{\alpha\mu\nu}$, the pure non-metricity sector remains completely free. The first condition is the same as obtained in New GR (see e.g. \cite{OrtinBook}), while the second condition comes from the mixed sector. It is remarkable that imposing the presence of this gauge symmetry precisely decouples $B_{\mu\nu}$ from the symmetric sector $h_{\mu\nu}$ and $H_{\mu\nu}$. 

Consistency of the symmetric sector requires the presence of yet additional gauge symmetries. We will impose having two copies of linearised Diffs so we have the gauge symmetry Diff$\times$Diff. As a matter of fact, since we know that the general theory already enjoys one Diff invariance, we only need to guarantee the presence of a second independent Diff. For that, we will impose the  Bianchi identity\footnote{The second Diff symmetry can be fully realised with either of the two symmetric fields and we have chosen to fully realise it with $h_{\mu\nu}$. Realising it with $H_{\mu\nu}$ would lead to the same results. This can be easily understood by considering the general transformations $\delta h_{\mu\nu}=\alpha_1\partial_{(\mu}\zeta^1_{\nu)}+\alpha_2\partial_{(\mu}\zeta^2_{\nu)}$ and $\delta H_{\mu\nu}=\beta_1\partial_{(\mu}\zeta^1_{\nu)}+\beta_2\partial_{(\mu}\zeta^2_{\nu)}$ that can be trivially diagonalised with a redefinition of the gauge parameters.}
\be
\partial_\mu\left(\frac{\delta \mS^{(2)}_\parallel}{\delta h_{\mu\nu}}\right)=0
\ee
that is identically satisfied if 
\be
\text{Diff}\times\text{Diff:}\quad\quad\quad c_5=2c_1,\quad c_3=-c_1,\quad c_{24}=-2c_1,\quad\text{and}\quad b_2=-b_1,
\ee
assuming that \eqref{eq:gaugeB} is already satisfied. If we further want $H_{\mu\nu}$ and $h_{\mu\nu}$ to decouple we need to impose the additional condition $b_1=4c_1$, in which case the quadratic Lagrangian reduces to
\be
\Lag^{(2)}=-c_1h_{\alpha\beta}\mathcal{E}^{\alpha\beta\mu\nu}h_{\mu\nu}-\frac{2 a_1 + a_2 - 4 c_1}{4}H_{\alpha\beta}\mathcal{E}^{\alpha\beta\mu\nu}H_{\mu\nu}+\frac{2a_1-a_2}{24}F^2
\ee
with $\mathcal{E}^{\alpha\beta\mu\nu}$ the Minkowski Lichnerowicz operator and $F_{\mu\nu\rho}=3\partial_{[\mu} B_{\nu\rho]}$ the Kalb-Ramond field strength. Requiring the decoupling of $h_{\mu\nu}$ and $H_{\mu\nu}$ guarantees the recovery of the appropriate Newtonian limit where only the metric perturbations couple to matter fields. Otherwise, via diagonalising the symmetric sector, we would generate an additional gravitational force for matter fields mediated by $H_{\mu\nu}$. It is interesting to notice however, that this additional force could be absorbed into a redefinition of Newton's constant at that order. This contrasts with other modified gravity scenarios where the diagonalisation involves a scalar field that only couples to the trace of the energy-momentum tensor and, therefore, can be detected by comparing the Newtonian and the lensing potentials. We finally notice that the three gauge symmetries that we have obtained amount to a full decoupling of the original Diffs for the three fields, i.e., the transformations \eqref{eq:Diff1} and $\eqref{eq:Diff2}$ become symmetries with independent parameters for $h_{\mu\nu}$, $H_{\mu\nu}$ and $B_{\mu\nu}$.

Another way of guaranteeing the propagation of 2 massless spin-2 fields is to complete the linearised Diffs to Diff$\times$WTDiff, i.e., by imposing an additional Weyl Transverse Diffeomorphism (WTDiff) invariance. Let us first obtain the condition to have TDiffs, which we achieve by imposing invariance of the field equations under a Diff with $\partial_\mu\zeta^\mu=0$. Similarly to the above case with the full Diff invariance, we can impose the symmetry either on $h_{\mu\nu}$ or on $H_{\mu\nu}$ independently thanks to the already existing Diff symmetry and to the fact that TDiff is a subgroup of Diff. The resulting condition in both cases is
\be
\text{TDiff:}\quad\quad\quad2c_1+c_{24}=0.
\ee 
Now, instead of completing this symmetry to full Diffs as before, we will complete it to add a linear Weyl symmetry. This can be realised in different ways. Unlike for TDiffs, a Weyl transformation is not, in general, a subgroup of Diffs so we need to consider the transformation $\delta h_{\mu\nu}=w_h\varphi\eta_{\mu\nu}$ and $\delta H_{\mu\nu}=w_H\varphi\eta_{\mu\nu}$, where we have given different Weyl weights $w_h$ and $w_H$ to the two fields. The associated Bianchi identity will be
\be 
\eta_{\mu\nu}\left[w_h\frac{\delta}{\delta h_{\mu\nu}}+w_H\frac{\delta }{\delta H_{\mu\nu}}\right]\mS^{(2)}_\parallel=0.
\ee
This identity is fulfilled if the following equations are satisfied:
\begin{align}
2 (2 a_1 + a_2) w_H - 2 (c_{24} - 8 c_3  - c_5) (w_h - w_H) - 
 b_1 (2 w_h - 3 w_H) - b_2 (4 w_h - 7 w_H)&=0,\nonumber\\
 b_1 w_h + 2 (b_2 + c_{24}  + 2 c_5) (w_h - w_H) - (2 a_1 + a_2) w_H&=0,\nonumber\\
 2 (c_{24} - 8 c_3 - c_5) (w_h - w_H) - (b_1 + 3 b_2) w_H&=0,\nonumber\\
  2(c_{24} + 2 c_5) (w_h - w_H) + b_1 w_H&=0.
  \label{eq:WTDiff}
\end{align}
Particularly interesting realisations of the additional Weyl symmetry are\footnote{See \cite{Iosifidis:2018zwo} for a detailed analysis of the different realisations of conformal/scale/Weyl transformations within the affine framework.}:
\begin{itemize}
\item $w_H=0$: The Weyl symmetry is fully realised on $h_{\mu\nu}$ while the connection does not transform. The solution is then $b_1+2b_2=3c_{24}-16c_3=8c_3+3c_5=0$.
\item $w_H=w_h$: The Weyl symmetry is covariantly realised with the change in the metric perturbation accompanied by the corresponding change in the connection. In this case we find $2a_1+a_2=b_1=b_2=0$.
\end{itemize}
We will end our discussion on the enhancement of the gauge symmetries by recalling that a massless spin-2 particle can be described by the WTDiff quadratic Lagrangian (see e.g. \cite{Alvarez:2006uu})
\be
\frac{1}{\mpl^2}\Lag_{\rm WTDiff}=\frac18\partial_\alpha h_{\mu\nu} \partial^\alpha h^{\mu\nu}-\frac14\partial_\mu h^{\mu\alpha}\partial^\nu h_{\nu\alpha}+\frac18\partial_\mu h\partial_\nu h^{\mu\nu}-\frac{3}{64} (\partial h)^2,
\label{eq:WTdiffLag}
\ee
whose only difference with GR is the appearance of a cosmological constant as an integration constant, but is otherwise completely equivalent.  It is straightforward to see that the general teleparallel theory identically reproduces the above WTDiff quadratic Lagrangian for the parameters $a_i=b_i=0$ and $c_5=-\frac83 c_3=-\frac12c_{24}=c_1$, which obviously satisfy Eqs. \eqref{eq:WTDiff}, and in terms of the perturbation $\hat{h}_{\mu\nu}\equiv h_{\mu\nu}-H_{\mu\nu}$. These parameters hence provide an alternative class of theories with a minimal linear spectrum containing just a massless spin-2 field, in the WTDiff realisation, in addition to the GTEGR discussed above. There are some crucial differences however because, while the GTEGR is guaranteed to preserve the number of dof's in the linear spectrum at the non-linear order due to the local $\GL$ symmetry of the full theory, there is no reason {\it a priori} to expect that the WTDiff alternative theories with a minimal spectrum around Minkowski will not introduce additional dof's in the non-linear spectrum. These theories are natural candidates for a teleparallel equivalent of unimodular gravity, whose linearisation is precisely the WTDiff Lagrangian \eqref{eq:WTdiffLag}, but there are some indications that the equivalence cannot be maintained at the non-linear order. Firstly, as we have discussed above, imposing a local $\GL$ symmetry for the linear theory singles out the GTEGR up to the degeneracy between $c_2$ and $c_4$, so the candidate for the unimodular equivalent cannot realise a local $\GL$ symmetry associated to the trivialisation of the connection sector. We would then expect the appearance of dof's associated to the connection from the non-linear terms which would in turn be strongly coupled around Minkowski. A second obstruction that takes place already at linear order is that the WTDiff is obtained for the perturbation $\hat{h}_{\mu\nu}$, while matter fields will only couple to $h_{\mu\nu}$ so the symmetry between $h_{\mu\nu}$ and $H_{\mu\nu}$ is broken by the matter sector.

\subsection{Theories with local Lorentz invariance}
A class of theories with enhanced interesting symmetries are those with a local Lorentz invariance. As we explained above, the local Lorentz invariance, at linear order, is achieved by removing $B_{\mu\nu}$ from the Lagrangian or, equivalently, by choosing parameters such as to trivialise its equation of motion. It is interesting to note that, since only $H_{\mu\nu}$ contributes to the non-metricity to this order, only the terms involving the torsion are relevant concerning the realisation of a local Lorentz symmetry. In particular, this means that the coefficients $c_{i}$ are not constrained by this requirement. Remarkably, when requiring the existence of the local Lorentz symmetry we obtain $b_3=b_1$, $a_3=-4a_1$ and $a_2=2a_1$, i.e., the pure torsion sector must reduce to the TEGR parameters and the mixed sector leaves $b_2$ free and $b_3=b_1$. This of course comprises the GTEGR case discussed above.


\section{Discussion}

In this Letter we have clarified how the TEGR and the STEGR descriptions of GR arise from the general equivalent in a teleparallel geometry as two particularly interesting gauges where the connection is further required to be either metric-compatible or torsion-free. The singular nature of GTEGR has been revealed to be a complete gauging of the global $\GL$ symmetry enjoyed by the general inertial connection. Furthermore, we have shown that requiring such a gauging of the global symmetry in the linear theory around Minkowski only leaves one free parameter.

We have then obtained the quadratic Lagrangian around Minkowski for the general quadratic teleparallel theory and discussed the necessity of having additional symmetries in order to obtain physically sensible theories. A few cautionary comments concerning the theories with enhanced symmetries are in order however. Firstly, imposing the required gauge symmetries at linear order is a necessary but not a sufficient condition. For instance, in New GR, it has been observed that the gauge symmetry rendering the 2-form stable at linear order cannot be maintained at the non-linear order \cite{Jimenez:2019tkx}. The crucial point is that tuning the coefficients in a somewhat ad-hoc manner to obtain the desired symmetries for the linear theory is not a very stable procedure and the inclusion of the interactions typically reveals the accidental nature of such gauge symmetries, thus signalling a discontinuity in the number of dof's which will be at the heart of the strong coupling problems coming in at an offending low scale.

The second cautionary remark regarding the additional symmetric rank-2 field is that even if the additional massless spin-2 field can be made to enjoy the necessary symmetries at linear order, it is expected that at fully nonlinear order the two spin-2 fields will interact, but it is known that a theory with massless spin-2 fields in its spectrum only admits one single species of this type. Another possibility one could envision is that one of the spin-2 fields becomes massive with healthy interactions. However, also in this situation the two spin-2 fields would present derivative interactions which are prone to the re-introduction of pathological modes \cite{deRham:2013tfa,deRham:2015rxa}. The potentially safe self-interactions obtained in \cite{Hinterbichler:2013eza} do not seem to be possible to realise in general, but only in very specific theories and, at most, up to some order in perturbation theory.  We will leave this for future work, but it is likely that the only consistent theory among the general quadratic teleparallel theories that includes gravity is in turn the GTEGR so no extensions would be possible along this path.

To finalise our cautionary discussion, we comment on a somewhat extended {\it folk} argument in favor of teleparallel theories claiming that they provide a better starting point for modifications of gravity. The alluded reason is that the action only contains first order derivatives of the fields and, consequently, the corresponding extensions are less prone to introducing Ostrogradski instabilities than the curvature based theories that contain second derivatives of the metric. This reasoning is not correct\footnote{A similar argument was also used to argue that higher order curvature theories in the metric-affine formalism avoided the instabilities thanks to the second order nature of the field equations. However, it has been shown that the metric-affine formulation of those theories are also generally plagued by ghosts \cite{BeltranJimenez:2019acz}.} and, actually, the crucial property is the introduction of additional dof's associated with the loss of local symmetries. Instead, we could use the Einstein action for GR (see footnote \ref{EinsteinLag}) where only first derivatives appear and, therefore, the starting point would also be an action with only first derivatives, very  much like the teleparallel framework. However, since the Einstein Lagrangian is not an exact Diff-scalar, utilising it as starting point to explore modifications will lead to the loss of symmetries and the problems will reappear as a consequence of the breaking of Diffs. The very same difficulties occur for the teleparallel  theories. The equivalents of GR crucially realise some symmetries up to total derivatives so one must be very careful when considering either extensions or non-standard matter couplings in order not to introduce additional potentially unstable dof's.
\\

{\bf Acknowledgements}:  JBJ acknowledges support from the  {\it Atracci\'on del Talento Cient\'ifico en Salamanca} programme and the MINECO's projects FIS2014-52837-P and FIS2016-78859-P (AEI/FEDER). LH is supported by funding from the European Research Council (ERC) under the European Unions Horizon 2020 research and innovation programme grant agreement No 801781 and by the Swiss National Science Foundation grant 179740. TSK was funded by the Estonian Research Council PRF project PRG356 and by the European Regional Development
Fund CoE TK133. AJC is supported by a PhD contract of the program FPU 2015 with reference
FPU15/02864 (Spanish Ministry of Economy and Competitiveness). This article is based upon work from CANTATA COST (European Cooperation in Science and Technology) action CA15117, EU Framework Programme Horizon 2020.

\end{document}